\documentclass[11pt,a4paper]{article}
\usepackage[english]{babel}
\usepackage{amsmath,amsthm,amssymb,epsfig,latexsym,color}
\usepackage[numbers,square,sort&compress]{natbib}
\usepackage{color}

\newcommand{\be}[1]{\begin{equation}\label{#1}}
\newcommand{\ee}{\end{equation}}
\newcommand{\bu}{\bar u}
\newcommand{\bv}{\bar v}
\newcommand{\bw}{\bar w}

\newcommand{\rank}{\mathop{\rm rank}}

\newcommand{\tr}{\mathop{\rm tr}}
\newcommand{\BB}[1]{|\Psi(#1)\rangle}
\newcommand{\CC}[1]{\langle\Psi(#1)}

\newcommand{\vk}{\kappa}
\newcommand{\tvk}{\widetilde{\kappa}}

\newtheorem{prop}{Proposition}[section]
 \makeatletter
 \@addtoreset{equation}{section}
 \makeatother

\begin{document}

\vspace{4pt}

\begin{center}
\begin{LARGE}
{\bf Why scalar products in the algebraic Bethe ansatz \\[14pt] have determinant representation}
\end{LARGE}

\vspace{44pt}

\begin{large}
{S. Belliard${}^\dagger$ and N.~A.~Slavnov${}^\ddagger$\   \footnote{samuel.belliard@lmpt.univ-tours.fr, nslavnov@mi-ras.ru}}
\end{large}

 \vspace{6mm}

  \vspace{4mm}

${}^\dagger$ {\it Institut Denis-Poisson, Universit\'e de Tours, Universit\'e d'Orl\'eans, Parc de Grammont, 37200 Tours, FRANCE}

 \vspace{4mm}

${}^\ddagger$  {\it Steklov Mathematical Institute of Russian Academy of Sciences,\\ 8 Gubkina str., Moscow, 119991,  RUSSIA}

\end{center}

\vspace{10mm}

\begin{abstract}
We show that the scalar products of on-shell and off-shell Bethe vectors in the algebra1ic Bethe ansatz solvable models
satisfy a system of linear equations. We find solutions to this system for a wide class of integrable
models. We also apply our method to the XXX spin chain with
broken $U(1)$ symmetry.
\end{abstract}

\vspace{10mm}

\section{Introduction}

The algebraic Bethe ansatz (ABA) is a part of the Quantum Inverse Scattering Method developed by the Leningrad school \cite{FadST79,FadT79,FadLH96}.
This method allows us to effectively find the spectra of quantum integrable models. Besides, the ABA can be used for the study of correlation functions
\cite{BogIK93L,KitMT00,KitKMST12,GohKS04}. To solve the latter problem, it is necessary to have compact, convenient representations for scalar products of Bethe vectors.
To date, such representations are considered determinant formulas for the scalar products containing on-shell Bethe vectors (that is, eigenvectors of the transfer matrix).

The first formula of this type was conjectured by M. Gaudin \cite{Gau72,Gaud83} for the norm of on-shell Bethe vectors in the model
of one-dimensional bosons. This conjecture was proved in \cite{Kor82} for a wide class of integrable models within the framework of the
ABA approach. A determinant formula for the scalar product of on-shell and twisted on-shell Bethe vectors was obtained in \cite{KirS87}.
This result was generalized to the scalar product of on-shell and off-shell Bethe vectors  in \cite{Sla89}.  Later, the authors of \cite{KitMT99}
paid attention that the determinant for the scalar product of on-shell and off-shell Bethe vectors had the form of the Jacobian of the transfer matrix
eigenvalue.
Similar results were later obtained for the systems with open boundaries \cite{Wang03,KitKMNST07}
and for the systems with broken $U(1)$ symmetry \cite{BelP15,BelP15a,BelS19}.

In all the cases listed above, the Jacobian of the transfer matrix eigenvalue appeared. However, there was no explanation
to this fact. In the present paper, we give such the explanation. Namely, we show that the scalar product of on-shell and off-shell Bethe vectors
satisfies a homogeneous system of linear equations. Being a solution to this system, the scalar product must have a determinant representation.
The Jacobian of the transfer matrix eigenvalue arises due to special properties of this eigenvalue.

The method that we present here works for integrable models with rational and trigonometric $R$-matrices. Perhaps, it can be generalized to
the elliptic case as well. One can also consider models with periodic boundary conditions or models with open boundaries. The key role in our method
is played by the action of the transfer matrix on off-shell Bethe vector.

The paper is organized as follows. We derive a system of linear equations for the scalar products in section~\ref{S-SLE}.
This is the main result of our paper. In  sections~\ref{S-S} and~\ref{S-AS} we respectively discuss the questions of solvability
and uniqueness of the solution of the obtained system. In section~\ref{S-XXX} we apply our method to the XXX spin chain with
broken $U(1)$ symmetry. Several auxiliary formulas are gathered in appendix~\ref{A-SS}.

\section{System of linear equations\label{S-SLE}}

In this section, we describe a new method for computing scalar products. We assume,
that the reader is familiar with the basic concepts of the ABA. Otherwise, we refer him to the works
\cite{FadST79,FadT79,FadLH96,BogIK93L,Sla18,Skl88}.

The central object of the ABA is the transfer matrix operator $\mathcal{T}(z)$. It gives rise to the Hamiltonian of the
quantum system  and integrals of motion. The transfer matrix depends on a complex variable
$z$ and acts in some Hilbert space $\mathcal{H}$. The eigenvectors of this operator $|\Psi\rangle$  are called
on-shell Bethe vectors. They are also eigenvectors of the Hamiltonian of the quantum system under consideration. The on-shell
Bethe vectors are parameterized by sets of complex numbers $|\Psi\rangle=\BB{u_1,\dots,u_n}$, ($n=0,1,\dots$), which must satisfy
a system of equations (Bethe equations). Otherwise, if the parameters $\{u_1,\dots,u_n\}$ are arbitrary complex numbers,
we are dealing with off-shell vector Bethe.

To calculate scalar products, it is also necessary to introduce the notion of dual Bethe vectors $\langle\Psi|$. They belong to the dual
space $\mathcal{H}^*$ and are also parameterized by sets of complex numbers $\langle\Psi|=\CC{v_1,\dots,v_n}|$, ($n=0,1,\dots$). If these parameters
satisfy the Bethe equations, then such a vector is called the dual on-shell Bethe vector. It is the eigenvector of the transfer matrix
in the dual space. If no conditions are imposed on the parameters $\{v_1,\dots,v_n\}$, then this vector is called
dual off-shell Bethe vector.

We now proceed to the derivation of a system of linear equations for scalar products of dual on-shell Bethe vectors and off-shell Bethe vectors.
Let $\bu=\{u_1,\dots,u_{n+1}\}$ be a set of arbitrary complex numbers. Denote by $\bu_j$ the subsets of this set that are complementary to the elements
$u_j$ ($j=1,\dots,n+1$), that is $\bu_j=\bu\setminus u_j$. Then we can construct $n+1$  off-shell Bethe vectors $\BB{\bu_j}$.

Let a set $\bv=\{v_1,\dots,v_n\}$ satisfies the Bethe equations. Then a vector $\CC{\bv}|$ is a dual on-shell Bethe vector. Hence,
\be{actdu}
\CC{\bv}|\mathcal{T}(z)=\Lambda(z|\bv)\CC{\bv}|,
\ee
where $\Lambda(z|\bv)$ is the eigenvalue of the operator $\mathcal{T}(z)$ on the dual on-shell vector $\CC{\bv}|$.

Let us introduce $n+1$ scalar products of the dual on shell Bethe vector $\CC{\bv}|$ with off-shell Bethe vectors $\BB{\bu_j}$
\be{defXj}
X_j=\CC{\bv}\BB{\bu_j}.
\ee
It is easy to show that the scalar products $X_1,\dots,X_{n+1}$ satisfy a system of linear equations. For this we consider the
following expectation values:
\be{exp-val}
\CC{\bv}|\mathcal{T}(u_j)\BB{\bu_j},\qquad j=1,\dots,n+1.
\ee
They can be computed in two different ways. First, we can act with $\mathcal{T}(u_j)$ to the left via \eqref{actdu}
\be{actleft}
\CC{\bv}|\mathcal{T}(u_j)\BB{\bu_j}=\Lambda(u_j|\bv)\CC{\bv}\BB{\bu_j}=\Lambda(u_j|\bv)X_j.
\ee
On the other hand, we can act with $\mathcal{T}(u_j)$ to the right. The action of the transfer matrix on off-shell
Bethe vectors is the key formula of the ABA. For a wide class of the ABA solvable models it can be written as
\cite{FadST79,FadT79,FadLH96,BogIK93L,Skl88,BelC13,Bel15}
\be{actright}
\mathcal{T}(u_j)\BB{\bu_j}=\sum_{k=1}^{n+1}L_{jk}\BB{\bu_k},
\ee
where $L_{jk}$ are numerical coefficients. For the moment we do not specify them, however, we will consider specific examples later.
Then we obtain
\be{actright1}
\CC{\bv}|\mathcal{T}(u_j)\BB{\bu_j}=\sum_{k=1}^{n+1}L_{jk}\CC{\bv}\BB{\bu_k}=\sum_{k=1}^{n+1}L_{jk}X_k.
\ee
Comparing \eqref{actleft} and \eqref{actright1} we arrive at a system of linear equations for $X_j$
\be{LSE}
\sum_{k=1}^{n+1}M_{jk}X_k=0, \qquad k=1,\dots,n+1,
\ee
where
\be{Mkj0}
M_{jk}=L_{jk}-\delta_{jk}\Lambda(u_{j}|\bv), \qquad j,k=1,\dots,n+1.
\ee
This is the system of linear equations that we were looking for.

Observe that deriving this system we only used the action formula \eqref{actright}.
Thus, in any model in which the transfer matrix  action on the off-shell Bethe vector is given by equation \eqref{actright},
the scalar product of off-shell and on-shell Bethe vectors satisfies a homogeneous system of linear equations \eqref{LSE}. These models include models with rational
and trigonometric $R$-matrices. We can also consider both systems with periodic boundary conditions \cite{FadST79,FadT79} and open systems \cite{Skl88}.
The formula \eqref{actright} also remains true in cases where the monodromy matrix does not have a vacuum vector, and the traditional ABA
requires some modification \cite{BelC13,Bel15}.

Since the system \eqref{LSE} is homogeneous, its solutions (if any) are given by minors of the matrix $M$.
Therefore, the scalar product of off-shell and on-shell Bethe vectors must have a determinant representation.
On the other hand, due to the homogeneity of the system \eqref{LSE}, its solutions $X_j$ are determined up to a common factor.
Below we show that this factor can be found if we put arbitrary complex parameters $\bu$ to some specific
values for which the scalar product is known from traditional methods. As for the solvability of the system \eqref{LSE}, it is clear,
that in realistic systems it is solvable, since scalar products exist and, generally speaking, are non-vanishing. We will show,
that the solvability of the system \eqref{LSE} is closely related to the  properties of the transfer matrix eigenvalues $\Lambda(z|\bv)$.

\section{Solvability\label{S-S}}

Solvability of the system \eqref{LSE} depends on the transfer matrix eigenvalue $\Lambda(z|\bv)$ and the coefficients $L_{jk}$.
In their turn, the latter depend on the concrete model. In this section, we consider a class of models with periodic boundary conditions and
possessing a rational $R$-matrix
\be{Rmat}
R(u,v)=I+g(u,v)P,\qquad\qquad g(u,v)=\frac{c}{u-v}.
\ee
Here $I$ is the identity matrix, $P$ is the permutation matrix, $c$ is a constant. We choose such models for reasons of simplicity.
However, all the results of this section can be easily generalized to the
case of a trigonometric $R$-matrix and for the models with open boundaries.

\subsection{Notation and conventions \label{SS-NC}}

Recall that we denote sets of variables by a bar: $\bv=\{v_1,\dots,v_n\}$, $\bu=\{u_1,\dots,u_{n+1}\}$ and so on. A special notation
$\bu_k=\bu\setminus u_k$, $\bv_j=\bv\setminus v_j$ is used for the subsets complementary to the elements $u_k$, $v_j$ and so on.

To make formulas more compact, we use a shorthand notation for the products of the functions $g(u,v)$ \eqref{Rmat}.
Namely, if this function  depends on a (sub)set
of variables, then one should take a product with respect to the corresponding (sub)set. For example,
\be{shn}
g(u,\bv)=\prod_{v_i\in\bv}g(u,v_i),\qquad g(u_k,\bu_k)=\prod_{\substack{u_i\in\bu\\u_i\ne u_k}}g(u_k,u_i)\qquad\text{and so on.}
\ee
By definition any product over the empty set is equal to $1$.

Finally, we also introduce
\be{DD}
\Delta(\bu)=\prod_{\substack{u_j,u_k\in\bu\\j>k}}g(u_j,u_k)\qquad\qquad \Delta'(\bu)=\prod_{\substack{u_j,u_k\in\bu\\j<k}}g(u_j,u_k).
\ee
These products naturally appear in the prefactors of determinant representations.

\subsection{Transfer matrix eigenvalue\label{SS-TME}}

In the models with rational $R$-matrix, the transfer matrix eigenvalue has the form
\be{LY}
\Lambda(z|\bv)=g(z,\bv)\mathcal{Y}(z|\bv),
\ee
where $\mathcal{Y}(z|\bv)$ is symmetric over $\bv$ (due to the symmetry of the Bethe vector) and linearly depends on every $v_j$. Thus,
\be{Y}
\mathcal{Y}(z|\bv)=\sum_{p=0}^{n}\alpha_p(z)\sigma^{(n)}_p(\bv).
\ee
Here $\alpha_p(z)$ are free functional parameters, and the functions $\sigma^{(n)}_p(\bv)$ are elementary symmetric polynomials in $\bv$:
\be{ESP}
\begin{aligned}
&\sigma^{(n)}_p(\bv)=\frac1{(n-p)!}\frac{d^{n-p}}{dt^{n-p}}\prod_{i=1}^n(t+v_i)\Bigr|_{t=0}, \qquad p=0,\dots,n,\\
&\sigma^{(n)}_p(\bv)=0, \qquad p<0,\quad\text{or}\quad p>n.
\end{aligned}
\ee
The superscript $n$ denotes the cardinality of the set $\bv$. The subscript $p$ is equal to the total degree of the polynomial.

In particular, a typical form of the $\mathcal{Y}$-function in the ABA solvable models is \cite{FadST79,FadT79}
\be{Y-BASM}
\mathcal{Y}(z|\bv)=\lambda_1(z)\prod_{i=1}^n\frac{z-v_i-c}c+\lambda_2(z)\prod_{i=1}^n\frac{z-v_i+c}c,  
\ee
where $\lambda_1(z)$ and $\lambda_2(z)$ are vacuum eigenvalues of the diagonal monodromy matrix entries. Obviously, this
$\mathcal{Y}$-function is the particular case of \eqref{Y}. The case of the modified algebraic Bethe ansatz \cite{BelC13,Bel15,Cram14,BP152,ABGP15}  is also
described by \eqref{Y} (see section~\ref{S-XXX}).

The coefficients $L_{jk}$ are related to the residues of the transfer matrix eigenvalue \cite{FadST79,FadT79}:
\be{LjkL}
L_{jk}=g(u_k,\bu_k) \mathcal{Y}(u_k|\bu_j).
\ee
It is easy to see from the action formula \eqref{actright} that if $L_{jk}=0$ for $k\ne j$, then the corresponding vector becomes on-shell Bethe vector.
On the other hand, this requirement leads to the absence of the poles of the transfer matrix eigenvalue $\Lambda(u_j|\bu_j)$ at $u_j=u_k$.

Thus, for this class of models, the matrix $M$ \eqref{Mkj0} takes the form
\be{Mkj}
M_{jk}=g(u_k,\bu_k) \mathcal{Y}(u_k|\bu_j)-\delta_{jk}\Lambda(u_{j}|\bv), \qquad j,k=1,\dots,n+1.
\ee

\subsection{Transformation of the system\label{SS-TS}}

The system of equations \eqref{LSE} has a non-trivial solution, if $\det M=0$. Let us prove that the determinant of the
matrix $M$ \eqref{Mkj} does vanish.
For this, we introduce an $n\times (n+1)$ matrix $\Omega_{jk}$ with the elements
\be{Sjk}
\Omega_{jk}=\frac{c}{g(u_k,\bv)}\frac{\partial \Lambda(u_k|\bv)}{\partial v_j}, \qquad j=1,\dots,n,\quad k=1,\dots,n+1.
\ee
One can easily show that
\be{Sjk1}
\Omega_{jk}=g(u_k,v_j)\mathcal{Y}(u_k|\{u_k,\bv_j\}).
\ee
Indeed, substituting \eqref{LY} into definition \eqref{Sjk} we obtain
\be{Sjk2}
\Omega_{jk}=g(u_k,v_j)\mathcal{Y}(u_k|\bv)+c\frac{\partial \mathcal{Y}(u_k|\bv)}{\partial v_j}.
\ee
Using the following properties of the elementary symmetric polynomials
\be{sigs}
\sigma^{(n)}_p(\bv)=v_j\sigma^{(n-1)}_{p-1}(\bv_j)+\sigma^{(n-1)}_p(\bv_j), \qquad
\frac{\partial \sigma^{(n)}_p(\bv)}{\partial v_j}=\sigma^{(n-1)}_{p-1}(\bv_j),
\ee
we immediately arrive at \eqref{Sjk1}

Let $\hat\Omega_\ell$ denote an $n\times n$ minor of this matrix in which the $\ell$th column is excluded:
\be{HS}
\hat\Omega_\ell=\det_{k\ne\ell} \Omega_{jk}.
\ee

\begin{prop}\label{P-SolSys}
Let the function $\mathcal{Y}(z|\bv)$ be given by \eqref{Y}. Then $\det M=0$, where $M$ is given by \eqref{Mkj}.
If $\rank{M}=n$, then the solution to the system \eqref{LSE} is
\be{SolX}
X_\ell=\phi(\bv)\Delta(\bu_\ell)\hat\Omega_\ell,
\ee
where $\phi(\bv)$ is a function of the parameters $\bv$.
\end{prop}

{\sl Proof.} Let $\widetilde{M}=WM$, where $W$ is a non-degenerated matrix. Then the original system $MX=0$ \eqref{LSE}
is equivalent to a new system $\widetilde{M}X=0$.

Consider an $(n+1)\times(n+1)$ matrix
\be{W1}
W_{jk}=g(u_k,w_j)\frac{g(u_k,\bu_k)}{g(u_k,\bw)}, \qquad\text{with}\qquad \det W=\frac{\Delta(\bu)}{\Delta(\bw)}.
\ee
Here $\bw=\{w_1,\dots,w_{n+1}\}$ is a set of generic pair-wise distinct complex numbers. Thus, the matrix
$W$ is not degenerated.

Then
\be{wtM}
\widetilde{M}_{jk}=g(u_k,\bu_k)\sum_{\ell=1}^{n+1}g(u_\ell,\bu_\ell)\mathcal{Y}(u_k|\bu_\ell)
\frac{g(u_\ell,w_j)}{g(u_\ell,\bw)} - g(u_k,w_j)\frac{g(u_k,\bu_k)}{g(u_k,\bw)}\Lambda(u_k|\bv).
\ee
Straightforward calculation shows (see appendix~\ref{A-SS} for details) that
\be{wtM1}
\widetilde{M}_{jk}=g(u_k,\bu_k)\left\{\mathcal{Y}(u_k|\bw_j) - \frac{g(u_k,w_j)}{g(u_k,\bw)}\Lambda(u_k|\bv)\right\}.
\ee
We now set $w_j=v_j$ for $j=1,\dots,n$.
Then the last row of $\widetilde{M}$ vanishes:
\be{wtMn1}
\widetilde{M}_{n+1,k}=g(u_k,\bu_k)\left\{\mathcal{Y}(u_k|\bv) - \frac{1}{g(u_k,\bv)}\Lambda(u_k|\bv)\right\}=0.
\ee
Thus, $\det\widetilde{M}=0$, and hence, $\det M=0$. The first statement of proposition~\ref{P-SolSys} is proved.

In all other rows of $\widetilde{M}$ with $j\le n$, we obtain
\be{wtMjk}
\widetilde{M}_{jk}=g(u_k,\bu_k)\left\{\mathcal{Y}(u_k|\{\bv_j,w_{n+1}\}) - \frac{g(u_k,v_j)}{g(u_k,w_{n+1})}\mathcal{Y}(u_k|\bv)\right\}, \qquad j\le n,
\ee
and using \eqref{ESP} we arrive at
\be{wtMjk1}
\widetilde{M}_{jk}=\frac{g(u_k,\bu_k)}{g(w_{n+1},v_j)}\Omega_{jk}, \qquad j\le n.
\ee
Thus, the original system \eqref{LSE} is equivalent to a system
\be{LSE01}
\sum_{k=1}^{n+1}g(u_k,\bu_k)\Omega_{jk}X_k=0, \qquad j=1,\dots,n.
\ee

If $\rank(M)=n$, then $\rank(\widetilde{M})=n$. Hence,  there exists some $m\in[1,\dots,n+1]$ such that $\hat\Omega_m\ne 0$.
 Then solving the system \eqref{LSE01} we obtain
\be{XlXm}
\frac{X_\ell}{\Delta(\bu_\ell)\hat\Omega_\ell}=\frac{X_m}{\Delta(\bu_m)\hat\Omega_m}
\ee
for all $\ell\in[1,\dots,n+1]$. The rhs of \eqref{XlXm} does not depend on $u_m$, but it depends on all other $\bu_m$.
Similarly, the lhs of \eqref{XlXm} does not depend on $u_\ell$, but it depends on all other $\bu_\ell$ for any $\ell\in[1,\dots,n+1]$.
We conclude that the ratios in the both sides of \eqref{XlXm} do not depend on the variables $\bu$. In this way we arrive at \eqref{SolX}. \qed

{\sl Remark.} Formally, $X_\ell$ given by \eqref{SolX} solve the system \eqref{LSE} for arbitrary $\phi(\bv)$. However, due to
the symmetry of the scalar product over $\bv$, the function $\phi(\bv)$  should be antisymmetric with respect to permutations in the set $\bv$.
Thus, we can present $X_\ell$ in the form
\be{SolX1}
X_\ell=\Phi(\bv)\Delta(\bu_\ell)\Delta'(\bv)\hat\Omega_\ell,
\ee
where $\Phi(\bv)$ is a symmetric function of the parameters $\bv$.

\section{Ambiguity of the solution\label{S-AS}}

Assuming that $\rank(M)=n$ we have found the solution to the system \eqref{LSE} up to the factor $\Phi(\bv)$.
How to find this factor? For this, we should consider a concrete model and choose the free parameters $\bu$ in such
a way, that the scalar product is known from other sources.

For example, let us take the inhomogeneous XXX chain with inhomogeneity parameters $\theta_1,\dots,\theta_N$ ($N$ being the length of the
chain). Then, if $\bu_{n+1}=\{u_1,\dots,u_n\}$ coincide with any subset of inhomogeneities of the cardinality $n$ (say,
$\bar\theta=\{\theta_1,\dots,\theta_n\}$), then the scalar product is equal to the partition function of the six-vertex model with domain wall
boundary condition \cite{Kor82}\footnote{%
Actually, this partition function is a particular case of the scalar product for all periodic models
solvable by the traditional ABA method.}. A determinant representation for this particular case of the scalar product was obtained
in \cite{Ize87}
\begin{multline}\label{SPIze}
\CC{\bv}\BB{\bar\theta}=c^{2n-n^2}\Delta(\bar\theta)\Delta'(\bv)\left(\prod_{a=1}^N\prod_{\mu=1}^n(\theta_\mu-\theta_a+c)(v_\mu-\theta_a)\right)\\
\times
\left(\prod_{\nu,\;\mu=1}^n(v_\mu-\theta_\nu+c)\right)\det\left(\frac{1}{(v_j-\theta_k)(v_j-\theta_k+c)}\right).
\end{multline}
On the other hand, specifying the solution \eqref{SolX1} to the case of the inhomogeneous XXX chain and comparing it with \eqref{SPIze} we
easily find that $\Phi(\bv)=1$. An example of a non-trivial function $\Phi(\bv)$ will be given in section~\ref{S-XXX}.

All the considerations above concerned the case $\rank(M)=n$. Is it possible that $\rank(M)<n$? The answer to this question depends
on the free functional parameters $\alpha_p(z)$ \eqref{Y}, or equivalently, on the concrete model under consideration. For example, a
function
\be{Ytr}
\mathcal{Y}(z|\bv)=\frac1{g(z,\bv)}
\ee
is a particular case of \eqref{Y}. Obviously, we have $\Lambda(z|\bv)=1$ in this case. This leads to $\Omega_{jk}=0$, and hence,
$\rank(M)=0$. Thus, even being restricted to the class of the $\mathcal{Y}$-functions \eqref{Y}, we formally might deal with the situation
$\rank(M)<n$. However, most probably that these cases correspond to trivial models which are not of physical interest. In particular, the
monodromy matrix in the case considered above is proportional to the identity operator. Obviously, all the scalar products in this model
vanish.

In the models of physical interest, the parameters $\alpha_p(z)$ are analytical functions of $z$. Therefore,
the minors $\hat\Omega_\ell$ of the matrix $\Omega$ are analytical functions of $\bu_\ell$. Thus,
in order to check the condition $\rank(M)=n$, one should again consider a concrete model and prove that these minors
are non-vanishing at least for some special choice of $\bu_\ell$.
For instance, we can send $\bu_{n+1}\to\bv$ in $\hat\Omega_{n+1}$. Then the matrix elements $\Omega_{jk}$ go to the entries of the Gaudin matrix:
\be{Gaud}
\Omega_{jk}\to \frac{\partial\mathcal{Y}(v_k|\bv)}{\partial v_j}, \qquad u_j\to v_j, \qquad j=1,\dots,n.
\ee
In many models this matrix is positively defined, and hence, its determinant cannot vanish identically.
Moreover, since the determinant of the Gaudin matrix describes the norm of the Hamiltonian eigenfunction, it cannot be zero.

\section{XXX chain with broken $U(1)$ symmetry\label{S-XXX}}

In previous sections, we have considered models with periodic boundary conditions and the rational $R$-matrix. We have shown that the reduction of the problem of calculating the scalar products to the system of linear equations enables us to easily reproduce the known results.  In this section, we consider an example of
the XXX chain with broken $U(1)$ symmetry. A modified algebraic Bethe ansatz (MABA) \cite{BelC13,Bel15} is used to solve this model (see also \cite{CYSW13a,CYSW13b,CYSW13c}).
The scalar products in this model were studied in the  works \cite{BelP15,BelP15a}. It was conjectured there that the scalar product of the on-shell and off-shell Bethe vectors is proportional to the Jacobian  of the transfer matrix eigenvalue. Recently this conjecture was proved for the spin-$1/2$ chains \cite{BelSV18sc,BelS19}. In this section, we prove this conjecture in the general case using our new method.

\subsection{MABA description of the model}

We give the basic notions of the MABA only. The reader can find more details in the papers \cite{BelC13,Bel15,BelSV18}.

We consider the XXX chain of the length $N$ with non-diagonal boundary conditions and with arbitrary positive (half)integer spins $s_i$ in the $i$th site.
Within the framework of the Quantum Inverse Scattering Method, this model can be obtained by a non-diagonal twist of a standard monodromy matrix
\be{TwMM}
T_K(u)=KT(u), \qquad K=\begin{pmatrix}\tvk&\vk_+\\\vk_-&\vk\end{pmatrix},
\ee
where $\vk$, $\tvk$, and $\vk_\pm$ are complex numbers. It is convenient to present the twist matrix $K$ in the form
$K=BDA$, where
\be{Mat-Tf}
A=\sqrt\mu\begin{pmatrix}1&\frac{\rho_2}{\vk^-}\\ \frac{\rho_1}{\vk^+}&1\end{pmatrix},\qquad
B=\sqrt\mu\begin{pmatrix} 1&\frac{\rho_1}{\vk^-}\\ \frac{\rho_2}{\vk^+}&1\end{pmatrix},\qquad
D=\begin{pmatrix} \tvk-\rho_1&0\\0& \vk-\rho_2\end{pmatrix},
\ee
and the parameters $\rho_i$ and $\mu$ enjoy the following constraints:
\be{murho}
\rho_1\rho_2-\rho_2\tvk-\rho_1\vk+\vk^+\vk^-=0, \qquad
\mu=\frac{1}{1-\frac{\rho_1\rho_2}{\vk^+\vk^-}}.
\ee
Then we have
\be{trTKa}
\mathcal{T}(u)=\tr \bigl(D\overline{T}(u)\bigr), \qquad  \overline{T}(u)=AT(u)B=
\begin{pmatrix} \nu_{11}(u)& \nu_{12}(u)\\ \nu_{21}(u)&\nu_{22}(u)\end{pmatrix}.
\ee
Thus, instead of the twisted monodromy matrix $T_K(u)$ we can consider a modified monodromy matrix
$\overline{T}(u)$ \eqref{trTKa}. Respectively, the transfer matrix $\mathcal{T}(u)=\tr T_K(u)$ now can be understood
as the trace of the twisted modified monodromy matrix $\overline{T}(u)$ with the diagonal twist $D$.

Off-shell (dual) Bethe vectors have the form
\be{Dos-ma}
\BB{\bu}=\prod_{j=1}^n \nu_{12}(u_j)|0\rangle, \qquad  \CC{\bu}|=\langle0|\prod_{j=1}^n \nu_{21}(u_j),\qquad n=0,1,\dots,S,
\ee
where $S=\sum_{i=1}^N 2s_i$. Here $|0\rangle$ and $\langle0|$ respectively are referent and dual referent states. These vectors
become on-shell, if $n=S$ and the parameters $\bu$ satisfy a system of Bethe equations (see \eqref{BEmod} below).

The transfer matrix eigenvalue $\Lambda(z|\bu)$ can be presented in the form \eqref{LY} with
\be{Ymod}
\mathcal{Y}(z|\bu)=(\tvk -\rho_1)\lambda_1(z)\prod_{i=1}^S(z-u_i-c)+(\vk -\rho_2)\lambda_2(z)\prod_{i=1}^S(z-u_i+c)+(\rho_1+\rho_2)F(z).
\ee
Here
\be{lam}
\lambda_1(z)=c^{-N}\prod_{i=1}^N\Bigl(z-\theta_i+c(s_i+\tfrac{1}{2})\Bigr), \qquad \lambda_2(z)=c^{-N}\prod_{i=1}^N\Bigl(z-\theta_i-c(s_i-\tfrac{1}{2})\Bigr),
\ee
where $\theta_i$ are inhomogeneities. The function $F(z)$ is given by
\be{F-rep}
F(z)=c^{-S-N}\prod_{i=1}^N\prod_{k=0}^{2s_i} \Bigl(z-\theta_i+c (s_i-k+\tfrac{1}{2})\Bigr).
\ee
Thus, the $\mathcal{Y}$-function \eqref{Ymod} belongs to the class \eqref{Y}, and the functions $\alpha_p(z)$ are polynomials in $z$
of degree $S+N$.

Finally, the system of Bethe equations reads
\be{BEmod}
\mathcal{Y}(u_j|\bu)=0, \qquad j=1,\dots,S.
\ee

\subsection{Scalar products in MABA}

Let $\CC{\bv}|$ with $\bv=\{v_1,\dots,v_S\}$ be a dual on-shell Bethe vector. Let $\bu=\{u_1,\dots,u_{S+1}\}$ be arbitrary complex numbers.
Consider  scalar products
\be{SPmaba}
X_{j}=\CC{\bv}\BB{\bu_j}, \qquad j=1,\dots,S+1,
\ee
of the dual on-shell Bethe vector $\CC{\bv}|$ with off-shell Bethe vectors $\BB{\bu_j}$.

All the technique described in the previous sections works in this case. First of all,
since the $\mathcal{Y}$-function \eqref{Ymod} belongs to the class \eqref{Y}, the system of linear equations \eqref{LSE} is solvable.
It remains to prove that $\rank(\Omega)=S$,  and then the solution is given by \eqref{SolX1}. For this, we prove that
$\hat\Omega_{\ell}\ne 0$ identically with respect to $\bu$. Obviously, without loss of generality it is enough to prove  that
$\hat\Omega_{S+1}\ne 0$ at least for $\bu\to\infty$. Indeed, since the $\mathcal{Y}$-function \eqref{Ymod} is  a polynomial in $z$,
the minors of the matrix $\Omega$ are polynomials in every $u_j\in\bu$. If these polynomials are non-vanishing at infinity, they
cannot be identically zero.

To consider the limit $\bu\to\infty$, it is convenient to transform the minor of the matrix $\Omega$ in which the last column is removed.
Let
\be{B}
B_{jk}=g(u_k,v_j)\frac{g(\bv_j,v_j)}{g(\bu,v_j)}, \qquad \det B=\frac{\Delta'(\bv)}{\Delta'(\bu_{S+1})}.
\ee
Then
\be{XB}
\hat\Omega_{S+1}=\frac{\Delta'(\bu_{S+1})}{\Delta'(\bv)}
\det\Bigl(\sum_{\ell=1}^Sg(u_j,v_\ell)\mathcal{Y}(u_j|\{u_j,v_\ell\})B_{\ell k}\Bigr).
\ee
The sum over $\ell$ is computed in appendix~\ref{A-SS} (see \eqref{H-XB}).
This leads us to
\be{XB1}
\Delta(\bu_{S+1})\Delta'(\bv)\hat\Omega_{S+1}=\det\left(\delta_{jk}\Lambda(u_j|\bv)+cg(u_j,\bu_j)\frac{\partial}{\partial u_k}\mathcal{Y}(t|\bu)\Bigr|_{t=u_j}\right).
\ee
Now we can set $u_k=Uk$ for $k=1,\dots,S$, and proceed to the limit  $U\to\infty$.
Straightforward calculation gives
\be{asyLL}
\Lambda(u_j|\bv)=\left(\tfrac{u_j}c\right)^N(\vk+\tvk) + o(U^{N}),\qquad U\to\infty,
\ee
and
\be{asyLY}
cg(u_j,\bu_j)\frac{\partial}{\partial u_k}\mathcal{Y}(t|\bu)\Bigr|_{t=u_j}=
\begin{cases}
&\left(\tfrac{u_j}c\right)^N(\rho_1+\rho_2-\vk-\tvk)+o(U^{N}),\qquad j=k,\\[11pt]
&o(U^{N}),\qquad j\ne k,
\end{cases}
\qquad U\to\infty.
\ee
Thus, we arrive at
\be{XB2}
\Delta(\bu_{S+1})\Delta'(\bv)\hat\Omega_{S+1}=(\rho_1+\rho_2)^S\left(\prod_{j=1}^S\left(\tfrac{u_j}c\right)^N\right)\bigl(1+o(1)\bigr),\qquad U\to\infty.
\ee
We see that $\hat\Omega_{S+1}\ne 0$ at $\bu\to\infty$. Hence, the scalar products $X_j$ are given by \eqref{SolX1}. It remains
to find the function $\Phi(\bv)$.

Using \eqref{Mat-Tf} and \eqref{trTKa} we obtain
\be{nu12}
\lim_{z\to\infty}\frac{\nu_{12}(z)}{z^N}=\frac\mu{\vk_-}(\rho_1+\rho_2)\mathbf{1}.
\ee
Then
\be{limSP}
\lim_{\bu\to\infty}\CC{\bv}\BB{\bu_{S+1}} \prod_{j=1}^S \left(\tfrac{u_j}c\right)^{-N} = \left(\frac\mu{\vk_-}(\rho_1+\rho_2)\right)^S\langle0|\prod_{j=1}^S\nu_{21}(v_j)|0\rangle.
\ee
Comparing this result with \eqref{SolX1} at $\bu\to\infty$ we conclude that
\be{Phires}
\Phi(\bv)=\left(\frac\mu{\vk_-}\right)^S\langle0|\prod_{j=1}^S\nu_{21}(v_j)|0\rangle.
\ee
Thus, we finally arrive at
\be{Xresmaba}
X_{S+1}=\left(\frac\mu{\vk_-}\right)^S\langle0|\prod_{j=1}^S\nu_{21}(v_j)|0\rangle \Delta(\bu_{S+1})\Delta'(\bv)\hat\Omega_{S+1},
\ee
and confirm the conjectures of \cite{BelP15}. It is worth mentioning that the expectation value in the prefactor of \eqref{Xresmaba} $\langle0|\prod_{j=1}^S\nu_{21}(v_j)|0\rangle$ also has a determinant representation. It is given
by a modified Izergin determinant \cite{BelSV18,BelS19}.

\section*{Conclusion}

We described a new method for computing the scalar products of Bethe vectors in the ABA solvable models.
We have shown that the scalar products of on-shell and off-shell Bethe vectors are described by a system of linear
equations. This explains the fact that these scalar products can be presented as determinants.

A key condition for the applicability of our method is a special form of the transfer matrix action on off-shell Bethe vectors.
Namely, this action should have the form \eqref{actright}. At the same time, explicit representation for the Bethe vector $\BB{\bu_j}$ is
not essential. The action \eqref{actright} immediately leads to the system of linear
equations \eqref{LSE} whose solutions are scalar products $\CC{\bv}\BB{\bu_j}$, where $\CC{\bv}|$
is on-shell Bethe vector.

We proved the solvability of the resulting system for a wide class of models with periodic boundary conditions and possessing the rational $R$-matrix.
However, generalization to the case of the trigonometric $R$-matrix and to the models with open boundaries is straightforward.  The question of uniqueness
of the solution should be solved separately in every concrete case. We have given several examples how to solve this problem.

In this paper we have restricted ourselves with reproducing the known results and proving some conjectures by the new method. We hope
however, that this simple method can also lead to new results. One of these directions is
the calculation of the scalar products in models with elliptic $R$-matrix. In this case, the action of the transfer matrix
does not have the form \eqref{actright}. However, our method apparently can be generalized to at least the case of
special values of the model parameters.

The most tempting direction is the study of scalar products in the models with symmetries of higher rank.
In these models, the Bethe vectors depend on several sets of variables, therefore, equation \eqref{actright} is replaced by a more complex
formula \cite{BelPRS13a,HutLPRS17b}. In particular, the result of the transfer matrix action on the off-shell Bethe vector contains terms in which the argument of the
transfer matrix replaces several Bethe parameters. The presence of such contributions is a serious obstacle for the application of our method to the models
with symmetries of higher rank.

It is not clear for today whether this problem is crucial or not. A series of results suggests that it can be solved. In particular, the transfer matrix action in models
with $\mathfrak{so}(3)$-symmetry contains similar contributions, however, a determinant representation for the  scalar product of
on-shell and off-shell vectors is known \cite{LiaPRS19}.
It is also worth mentioning that determinant formulas are known for some particular cases of the scalar products in models
with  $\mathfrak{gl}(3)$ and $\mathfrak{gl}(2|1)$ symmetry algebras \cite{BelPRS13,HutLPRS17a,Sla15}. This gives a hope that there is some
generalization of our method to the models with higher symmetry rank. Work in this direction is now underway.

\section*{Acknowledgements}
We thank R. A. Pimenta, X. Martin, S. Nicolis, and G. Lemarthe for useful discussions.
 N.S. thanks the CNRS for his grant to join the Institute Denis Poisson of the University of Tours where a part of this work was done.

\appendix

\section{Several sums\label{A-SS}}

Computing the matrix elements $\widetilde{M}_{jk}$ \eqref{wtM} we deal with a sum
\be{H-LA}
H_{jk}=\sum_{\ell=1}^{n+1}g(u_\ell,\bu_\ell)\mathcal{Y}(u_k|\bu_\ell)
\frac{g(u_\ell,w_j)}{g(u_\ell,\bw)}.
\ee
Substituting here $\mathcal{Y}(u_k|\bu_\ell)$ in terms of the elementary symmetric polynomials \eqref{ESP} we obtain
\be{H-LA1}
H_{jk}=\sum_{p=0}^n\frac{\alpha_p(u_k)}{(n-p)!}\frac{d^{n-p}}{dt^{n-p}}\prod_{\mu=1}^{n+1}(t+u_\mu)
\sum_{\ell=1}^{n+1}\frac{g(u_\ell,w_j)}{t+u_\ell}\frac{g(u_\ell,\bu_\ell)}{g(u_\ell,\bw)}\Bigr|_{t=0}.
\ee
Let
\be{G-LA1}
G=\sum_{\ell=1}^{n+1}\frac{g(u_\ell,w_j)}{t+u_\ell}\frac{g(u_\ell,\bu_\ell)}{g(u_\ell,\bw)}.
\ee
Consider a contour integral
\be{I-LA1}
I=\oint_{|z|=R\to\infty}\frac{dz}{2\pi ic}\frac{g(z,w_j)}{t+z}\frac{g(z,\bu)}{g(z,\bw)}.
\ee
Here the anticlockwise oriented contour is taken around infinity.
Obviously, $I=0$ as the integrand behaves as $z^{-2}$ at $z\to\infty$. On the other hand, the sum of the residues within the contour gives
\be{I-LA2}
I=G-\frac{1}{t+w_j}\prod_{\mu=1}^{n+1}\frac{t+w_\mu}{t+u_\mu}.
\ee
Hence,
\be{G-LA2}
G=\frac{1}{t+w_j}\prod_{\mu=1}^{n+1}\frac{t+w_\mu}{t+u_\mu}.
\ee
Substituting this into \eqref{H-LA1} we obtain
\be{H-LA2}
H_{jk}=\sum_{p=0}^n\frac{\alpha_p(u_k)}{(n-p)!}\frac{d^{n-p}}{dt^{n-p}}\frac{1}{t+w_j}\prod_{\mu=1}^{n+1}(t+w_\mu)\Bigr|_{t=0}=
\mathcal{Y}(u_k|\bw_j).
\ee

\vspace{5mm}

Now we compute the sum in \eqref{XB}. Let
\be{H-XB}
H_{jk}=\sum_{\ell=1}^Sg(u_j,v_\ell)\mathcal{Y}(u_j|\{u_j,v_\ell\})
g(u_k,v_\ell)\frac{g(\bv_\ell,v_\ell)}{g(\bu,v_\ell)}.
\ee
Then
\be{H-XB1}
H_{jk}=\sum_{p=0}^S\frac{\alpha_p(u_j)}{(S-p)!}\frac{d^{S-p}}{dw^{S-p}}\prod_{\mu=1}^S(w+v_\mu)
\sum_{\ell=1}^S g(u_j,v_\ell)g(u_k,v_\ell)\frac{g(\bv_\ell,v_\ell)}{g(\bu,v_\ell)}\frac{w+u_j}{w+v_\ell}\Bigr|_{w=0}.
\ee
Let
\be{G-XB1}
G=\sum_{\ell=1}^S g(u_j,v_\ell)g(u_k,v_\ell)\frac{g(\bv_\ell,v_\ell)}{g(\bu,v_\ell)}\frac{w+u_j}{w+v_\ell}.
\ee
Consider a contour integral
\be{I-XB1}
I=\oint_{|z|=R\to\infty}\frac{dz}{2\pi ic}g(u_j,z)g(u_k,z)\frac{g(\bv,z)}{g(\bu,z)}\frac{w+u_j}{w+z}.
\ee
Obviously, $I=0$. On the other hand, the sum of the residues within the contour gives
\be{I-XB2}
I=-G-\delta_{jk}\frac{g(\bv,u_j)}{g(\bu_j,u_j)}+\frac{c}{w+u_k}\prod_{\mu-1}^S\frac{w+u_\mu}{w+v_\mu}.
\ee
Thus,
\be{G-XB2}
G=\delta_{jk}\frac{g(u_j,\bv)}{g(u_j,\bu_j)}+\frac{c}{w+u_k}\prod_{\mu-1}^S\frac{w+u_\mu}{w+v_\mu}.
\ee
Substituting this into \eqref{H-XB1} we obtain
\begin{multline}\label{H-XB2}
H_{jk}=\delta_{jk}\mathcal{Y}(u_j|\bv)\frac{g(u_j,\bv)}{g(u_j,\bu_j)}+\sum_{p=0}^S\frac{\alpha_p(u_j)}{(S-p)!}\frac{d^{S-p}}{dw^{S-p}}\frac{c}{w+u_k}\prod_{\mu=1}^S(w+u_\mu)
\Bigr|_{w=0}\\
=\delta_{jk}\frac{\Lambda(u_j|\bv)}{g(u_j,\bu_j)}+c\frac{\partial}{\partial u_k}\sum_{p=0}^S\frac{\alpha_p(t)}{(S-p)!}\frac{d^{S-p}}{dw^{S-p}}\prod_{\mu=1}^S(w+u_\mu)
\Bigr|_{w=0,\;t=u_j}\\
=\delta_{jk}\frac{\Lambda(u_j|\bv)}{g(u_j,\bu_j)}+c\frac{\partial}{\partial u_k}\mathcal{Y}(t|\bu)\Bigr|_{t=u_j}.
\end{multline}

\end{document}